# Non-synchronism in Global Usage of Research Methods in Library and Information Science from 1990 to 2019

Chengzhi Zhang*, Liang Tian
Department of Information Management, Nanjing University of Science and Technology, Nanjing, 210094, China

**Abstract:** The global development of Library and Information Science (LIS) is influenced by various factors such as the economy, society, culture, discipline, tradition, and more. Consequently, the research methods of LIS vary greatly among countries. To better understand these differences, we conducted a study of 5,281 research papers from 81 countries published in internationally representative journals over the past thirty years. We manually annotated the research methods used in some articles through content analysis, and subsequently developed and trained a deep learning model for automatic classification of research methods. Using this method, we conducted a comparative analysis of the usage of research methods in different countries. Our findings reveal that there are differences in the research methods used across countries, with each country having its unique research profile and distribution of research methods. Even when investigating the same topic, research methods can differ between countries. Our study also uncovers that there are differences between the national and international distribution of research methods, these differences have decreased over the past 30 years. By highlighting the characteristics of discipline development in various countries from the perspective of research methods, our study can help guide discipline development at the national level. This study provides insights into the usage trends of research methods across different countries and highlights the unique characteristics of discipline development in each country. This information can be valuable in promoting collaboration and understanding between countries and in guiding discipline development at the national level.

## 1. Introduction

Library and Information Science (LIS) is an interdisciplinary field with research methods derived from different disciplines (Hjørland, 2005). There are concerns that the discipline may be losing its independence, necessitating self-analysis (Weller & Haider, 2007). Understanding research methods in LIS is essential for the development of teaching and research in the discipline (Järvelin & Vakkari, 1990). Scholars have investigated and analyzed research methods in LIS for decades, resulting in numerous studies (Blake, 1994; M. K. Rochester, 1995; Fidel, 2008; Riedl & Rueckel, 2011).

Qualitative methods have been utilized to explore the research methods available in the discipline (Chu & Ke, 2017; Järvelin & Vakkari, 1993), while quantitative methods have been used to analyze trends in their usage (Lund & Wang, 2021; Ma & Lund, 2021; Tuomaala et al., 2014). Research methods continue to be a recurring topic in LIS, with scholars emphasizing the importance of understanding and developing research methods to uphold the field's independence and enhance its teaching and research practices.

* Corresponding author: Chengzhi Zhang (zhangcz@njust.edu.cn).

The interdisciplinary nature of LIS requires a thorough understanding of the diverse research methods used within the field. Scholars have investigated and analyzed these research methods over time, emphasizing the need for self-reflection and progress in the discipline. Both qualitative and quantitative methods have been used to explore the research methods utilized in LIS. These studies remain pivotal in promoting the autonomy and enhancing the quality of research and education in the field.

While choosing an appropriate research method is crucial for any scientific study and is dependent on the research question, other factors, such as a country's economic, social, and cultural context, may also influence the choice of research methods. Objective constraints may lead to differences in research methods for the same research question in different countries. Previous studies have focused on analyzing general international trends (Ullah & Ameen, 2018; Ma & Lund, 2021) or a particular country or region (Lou *et al.*, 2021; Ullah & Ameen, 2021). Rochester & Vakkari (1998) explored the differences between national characteristics and international trends in research in the field of LIS more than two decades ago. However, with new research methods being developed, international differences and trends in the usage of research methods in the last two decades remain unclear.

This study aims to investigate the usage of national research methods in the field of LIS over the last 30 years (1990~2019) by examining critical academic journals in the discipline. The examination will provide insights into the usage and development of national research methods in LIS at the country level. To achieve this goal, the study aims to answer the following research questions:

***RQ1***: How have research methods in LIS been distributed across countries over the last three decades?

***RQ2***: How has the usage of research methods changed over time in each country over the last three decades?

***RQ3***: What are the characteristics of development research methods in LIS in different countries?

This paper provides insights into how national characteristics in LIS differ by examining the worldwide use of research methods from a country-level perspective. It offers a significant contribution to understanding the current state of research methods used in each country and highlights the characteristics of disciplinary development in each country from a research methods perspective. This guidance can help guide the development of the discipline at the national level.

## 2. Literature Review

This article focuses on research methods studies in LIS. This section provides a general review of research methods studies, followed by a summary of research methods studies in the field for various countries.

### 2.1 Research methods related to studies LIS

Schlatcher and Thomison (1974) conducted one of the earliest studies investigating research methods in LIS, focusing on the relationship between gender, degrees earned, institutions attended, years completed, and research methods used by dissertation authors. In 1990, Järvelin and Vakkari (1990) published influential work that generated many subsequent studies on research methods. Several studies have explored the usage of different research methods in LIS and identified trends (Hider & Pymm, 2008; M. K. Rochester, 1995; Tuomaala *et al.*, 2014).

The discussion of research methods in LIS has been multifaceted and detailed, covering various scopes and factors. Some studies have provided a detailed review of the usage and development of a particular research method or technique, such as the Delphi method (Lund, 2020) and technology acceptance models (F. Wang, & X. Wang, 2020). Others have analyzed more complex use of research methods, such as mixed research methods (Fidel, 2008; Granikov *et al.*, 2020). Additionally, the usage of research methods is discussed in relation to topics (Ma & Lund, 2021), authors (Lou *et al.*, 2021), and other factors.

Most previous studies have used manual coding-based content analysis methods. However, with the development of artificial intelligence techniques, scholars have started experimenting with automated methods to identify research methods used in academic papers (Eckle-Kohler *et al.*, 2013; Zhang, Tam & Cox, 2021; Zhang, Tian & Chu, 2021). In this paper, we adopt an automated approach to analyze research methods.

Since selecting and using research methods is a dynamic and changing process, the discussion of research methods in LIS has been an enduring topic for the past few decades. Understanding the usage of research methods is instructive for comprehending the development of a national discipline and the current state of research in a specific direction.

### 2.2 Research methods studies for different countries

Table 1 provides an overview of the studies on research methods in different countries. Many studies were conducted around the year 2000, where researchers reviewed and analyzed the use of research methods in LIS across different countries. However, we believe that several studies on the use of research methods may have been written in non-English languages or national journals, which lack an overall comparison of research methods usage worldwide over the past two decades. Therefore, analyzing the usage of research methods in different countries can help recognize international trends and regional differences in developing research methods in the discipline.

**Table 1. Summary of the literature on methodological studies of LIS in different countries**

| Authors | Countries surveyed | Period | # Documents |
|---|---|---|---|
| Rochester (1995) | Australia | 1985-1994 | 126 |
| Cheng (1996) | China | 1985,1990,1994 | 7042 |
| Ward (1997) | UK | 1965,1975,1985,1995 | 371 |
| Rochester & Vakkari (1998) | International (6 countries) | 1965-1995 | Compare studies from several different countries |
| Cano (1999) | Spain | 1977-1994 | 354 |
| Yontar & Yalvac (2000) | Turkey | 1952-1994 | 644 |
| Steinerová (2003) | Slovakia | 1990-2000 | - |
| Lou *et al.* (2021) | China | 1980-2020 | 2421 |
| Ullah & Ameen (2021) | Pakistani | 2001-2016 | 600 |
| Ngulube & Ukwoma (2021) | South Africa and Nigeria | 2009-2015 | 104 |

Rochester and Vakkari's study observed that the development of the LIS discipline varies from country to country due to factors such as the level of development of the discipline, economic level, culture, and tradition (Rochester & Vakkari, 1998). This study aims to reveal the differences in research methods usage in the LIS field across different countries by quantitatively analyzing academic papers published in international journals. By comparing the trends in research methods usage across countries over a long period, this study sheds light on the national-level development of the discipline, providing a better understanding of its nature and characteristics in different countries.

Compared to previous studies, this research uses automated methods to identify research methods used in academic papers, allowing for a broader range of data analysis and examination of continuous changes over long periods. Additionally, this study investigates all mainstream methods within LIS, rather than being limited to a specific research method. The survey covered more than 80 countries, providing a broader scope of analysis than previous studies.

## 3 Methodology

In this section, we first present the dataset and criteria for classifying research methods in Section 3.1. Subsequently, Section 3.2 describes the key techniques, including automatic research method classification and its results, as well as data analysis approach for examining the usage of research methods over countries.

### 3.1 Data source

We collected a total of 5,281 full-text research articles published from 1990 to 2019 in the Journal of the Association for Information Science and Technology (JASIST, n = 3,537), Journal of Documentation (JDoc, n = 1,102), and Library and Information Science Research (LISR, n = 642) to use as our data source. We also obtained the metadata of these articles, including the basic information of the authors.

This study focused on examining the application of research methods in the LIS field, so the required journals' topics should be closely related to LIS research. JASIST's topics focus on tools and techniques related to the information production, discovery, and documentation processes, etc. JDoc's topics focus on the presentation of original concepts or ideas, while LISR focuses on the research process in LIS, especially the demonstration and application of innovative methods and new theories. The topics of these three journals cover various aspects of LIS research, making them representative for exploring research methods in LIS. Previous related studies have also selected these journals (e.g. Chu, 2015; Chu & Ke, 2017; Ma & Lund, 2021), and choosing these journals in this study facilitates comparison with existing studies.

To reveal differences in the usage of research methods at the national level in the field of LIS, we need to identify the country of origin for each article. If an article has authors from multiple countries, we count it multiple times and attribute it to each relevant country. Our study examined a total of 5,281 research papers from 81 countries.

### 3.2 Coding scheme of research methods

In this article, we identified research methods of LIS research articles based on Table 2. Our coding scheme was adopted from Chu & Ke (2017), which considers data collection and analysis as two essential attributes of research methods. This coding scheme classifies and names research methods based on the data collection methods of research.

The training dataset for this study was obtained from Chu & Ke (2017). The blue bars in Figure 1 show the frequency distribution of research methods reported in the dataset from Chu & Ke (2017). However, some categories had too few cases, such as the Delphi study, as shown in the figure. Hence, we used a technique to expand each method category with less than 300 articles in the dataset. We chose the research method with less than 300 samples because these methods were less effective in classification during our pre-experiments. Our expansion technique consisted of four steps: 1) Use TextRank (Mihalcea & Tarau, 2004) to extract 20 keywords for each article in the existing dataset. 2) Select the top 10 keywords with the highest frequency for each method category with less than 300 articles. 3) Search Web of Science (http://webofscience.com/) using the chosen keywords for each category, and the abstracts of retrieval results are downloaded. 4) Manually filter and annotate the retrieval results and add the labeled data to the existing dataset. We used keywords instead of research method names or synonyms in this expansion process to obtain more data for those methods with fewer than 300 articles in each.

**Table 2. Coding scheme for research methods in this article**

| Research Method | Definitions |
| --- | --- |
| Bibliometrics | Bibliometric methods are used to collect information on publications and citations, and bibliometrics also includes citation analysis, informetrics, and scientometrics. |
| Content analysis | Content analysis refers to collecting data by conducting a systematic examination of texts or other passages in the contexts of their use, and content analysis also includes discourse analysis and secondary analysis. |
| Delphi study | The Delphi method is generally used to collect data through questionnaires from expert groups to address research questions reach a consensus and make predictions through multiple rounds of communication. |
| Ethnography/field study | Both can be applied when collecting data using a variety of techniques (e.g. observation and interviews) in the natural environment in which participants live or work. |
| Experiment | An experiment is an established method of data collection by following a procedure to test what is being studied in a laboratory or field environment. |
| Focus groups | Focus groups involve collecting data by discussing research questions between a facilitator and a group of participants. |
| Historical method | Historical methods refer to the collection of data by examining, synthesizing, summarizing, and interpreting existing published and unpublished material relevant to historical research questions. |
| Interview | The interview is a data collection technique in which individual participants are asked questions related to the research questions. |
| Observation | Observation is a method of collecting data by carefully and intently observing and recording the subject being studied. |
| Questionnaire | The questionnaire is a method of collecting data using a predefined list of questions. |
| Research diary/journal | A research diary/journal is a technique used to collect data about events, activities, thoughts, reflections, or other aspects of a diary collected by an individual who keeps a diary over some time. |
| Theoretical approach | A theoretical approach as a research method is a technique for collecting data through conceptual analysis, theoretical testing, or similar activities (e.g., conceptual analysis, modeling, & theory building). |
| Think aloud protocol | The thinking aloud protocol is a research method designed to collect data about participants' cognitive activity through the verbal presentation of their ideas. |
| Transaction log analysis | Transaction log analysis is used as a research method to collect data by analyzing transaction logs that are automatically captured on the server or client. |
| Webometrics[1] | Webometrics is defined as bibliometrics in a web environment, where web pages and websites are usually considered publications; internal links are |

[1] In this paper, we define webometrics as bibliometrics in a web environment. Following the approach of Chu and Ke (2017), we treat bibliometrics and Webometrics as distinct research methods. This approach is also adopted by Lou et al. (2021).

|  | considered citations, while external links are considered references. |
| --- | --- |
| Other methods | In addition to the 15 research methods mentioned above. |

As a result, we have obtained an expanded dataset, as shown by the orange bars in Figure 1. This expanded dataset will serve as the new training dataset for our current research and includes a total of 2,799 research articles. By including studies that use multiple research methods, we were able to increase the number of articles in certain method categories that previously had less than 300 articles. This approach has led to an increase in the number of articles in other method categories, such as experiments and questionnaires, as shown in Figure 1.

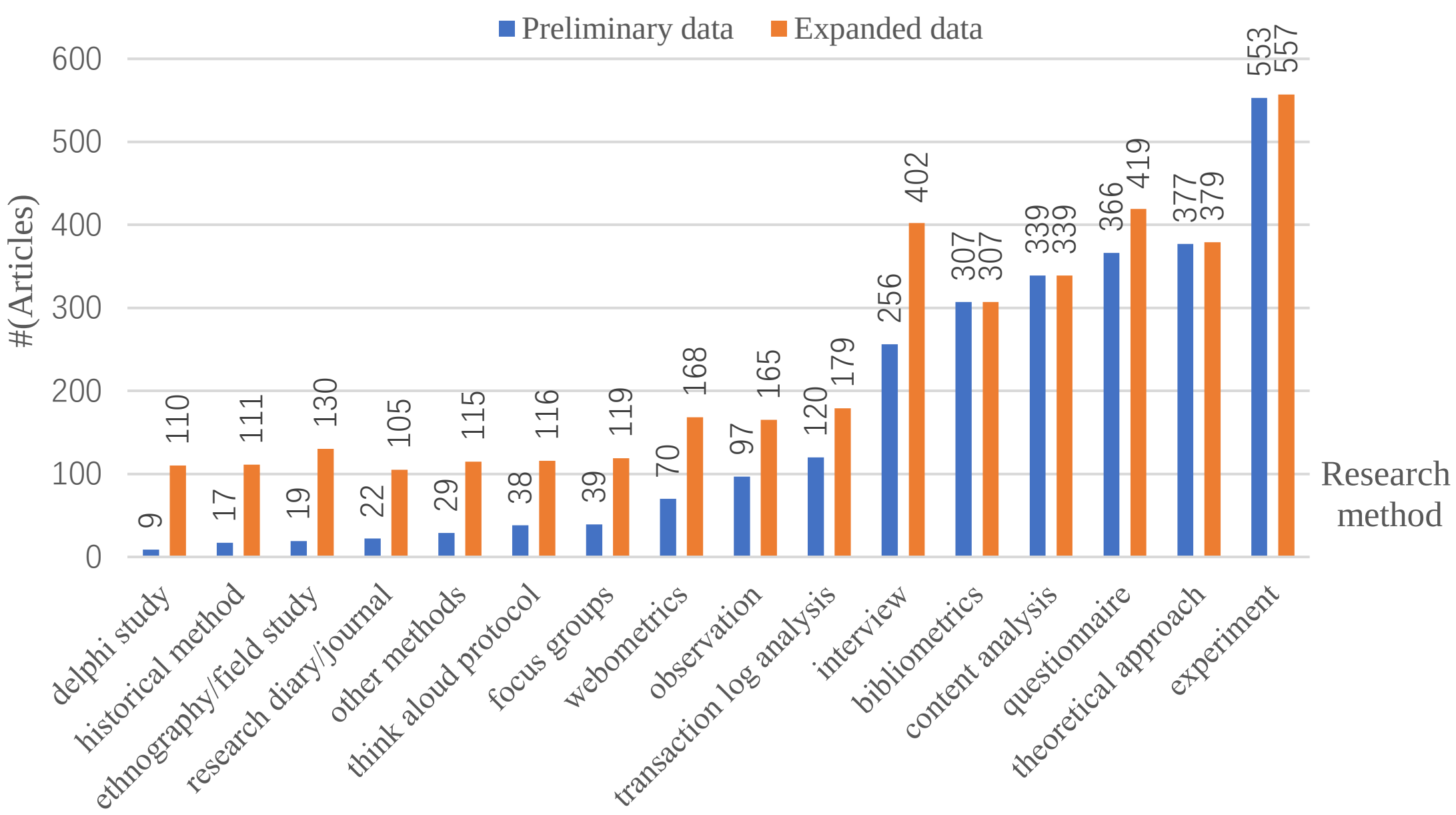

**Figure 1. Distribution of research methods in original and extended datasets**

### 3.3 Automatic classification of research methods

#### 3.3.1 Full-text cognitive-based classification model

In some cases, academic article abstracts may not provide a complete description of the research methods used, prompting the need for the full text to be used for automatic classification. However, the full text of academic articles can be lengthy and contain extraneous content that is unrelated to the research method. As a solution, we developed an automatic classification model for research methods based on full-text cognition (CogFT), inspired by CogLTX (Ding et al., 2020).

Our approach involves partitioning the full text of an article into multiple blocks, selecting the critical blocks related to research methods, and then classifying them. By leveraging the CogFT approach, we are able to extract relevant information from the full text of academic articles and improve the accuracy of automatic classification of research methods.

To facilitate the handling of full-text content, we divided it into 128 blocks, with each block having a length of 64. This division enables the classification model in our study to handle texts of lengths up to 8192. If the text is longer than 8192, we intercept the text from the front.

For input embedding, we used the pre-trained language model SciBERT (Beltagy et al., 2019) in the Summary Network. The objective of the Summary Network is to identify all descriptions in the full text related to the research method and minimize summary redundancy. We employed a global encoder based on Self-Attention to establish relationships between each block of the full text. Finally, through an output layer, the Summary Network obtained the probability of whether each block was key to the description of the research method. In this study, the top four blocks with the highest probability were merged and used as input to the Classification Network.

The Classification Network utilized SciBERT directly, where the summary obtained from the Summary Network was fed into SciBERT, and the output feature from SciBERT was classified using a Sigmoid layer at the end of the network.

To enable faster convergence, we utilized the same training algorithm as ConLTX (Ding et al., 2020) and initialized the summary. The summary initialization algorithm we employed was a combination of two classical unsupervised summary algorithms, MMR (Carbonell & Goldstein, 1998) and BM25 (Trotman et al., 2014). We ranked the 128 blocks using the scoring averages obtained by both algorithms.

In this study, the dataset was randomly divided into training, test, and validation sets in an 8:1:1 ratio. The Summary Network has two Self-Attention layers, each with eight attention heads, and a hidden layer size of 768. The model is optimized using Adam (Kingma & Ba, 2014) with a minimum batch size of 4 and gradient accumulation over four iterations due to GPU limitations. The learning rate used for training is 0.00001.

Datasets with noise are common in the field of artificial intelligence. For example, well-known datasets such as ImageNet and MS-COCO contain mislabeled data. Annotating research methods for academic papers requires extensive domain knowledge and is a difficult task. Therefore, even if strict annotation specifications are followed, the dataset may still have some annotation errors. In this section, we attempted to identify potential noisy data in the experimental dataset using Confident Learning (Northcutt et al., 2021) to estimate the noisy data in the dataset. We then explored the possibility of further improving the classification accuracy of the research method. A total of 809 potentially noisy data were identified from the expanded dataset using Confident Learning. We manually verified the potentially noisy data to determine which data were true labeling errors and updated the dataset based on the manual verification results. After updating the expanded dataset based on the results of manual verification of the potential noise data, the experiments were conducted again.

### 3.3.2 Classification results of research method

Since research method classification is a multi-label text classification task, evaluation metrics are divided into sample-based and label-based metrics (Zhang & Zhou, 2014). This study uses sample-based evaluation metrics (Samples_P, Samples_R, Samples_$F_1$) and label-based evaluation metrics (Micro_P, Micro_R, Micro_$F_1$, Macro_P, Macro_R, Macro_$F_1$) to evaluate the model's performance. Table 3 shows results for classifying research methods before/after confident learning. The cognitively-inspired full-text model (CogFT) for classifying research methods, used in this study, exhibited strong performance. The classification of research methods was enhanced after updating the expanded dataset, surpassing the performance prior to the update.

**Table 3. Results for classification of research methods**

| Model | Macro-P | Macro-R | Macro-$F_1$ | Micro-P | Micro-R | Micro-$F_1$ | Samples-P | Samples-R | Samples-$F_1$ |
|---|---|---|---|---|---|---|---|---|---|
| CogFT (without confident learning) | 0.8065 | 0.8129 | 0.8062 | 0.7802 | 0.7886 | 0.7844 | 0.8201 | 0.8379 | 0.8128 |
| CogFT (with confident learning) | 0.8266 | 0.8434 | 0.8299 | 0.8137 | 0.8204 | 0.8171 | 0.8756 | 0.8748 | 0.8551 |

### 3.4 Data analysis of the usage of research methods over countries

In this paper, our research questions aim to explore the distribution of research methods across different countries. We analyze the variations in research method distribution and changes over time to identify potential disciplinary developments that are specific to each country. To ensure comparability, we construct a vector of disciplinary research method distributions at the national level, which represents the frequency of usage of each research method in a country. By calculating the distance between these vectors, we can determine the similarity between research methods used in different countries.

We examine the utilization of research methods in each country in terms of the frequency of multiple method usage and the diversity of research methods utilized. An article is regarded as having used multiple methods if it employs more than one research method. We measure the diversity of research methods utilized in a country using the Shannon Variety Index (Spellerberg & Fedor, 2003).

## 4 Result

This section presents results of the study. Specifically, Section 4.1 offers descriptive statistics for publications originating from various countries. Section 4.2 presents overall distribution of research methods used in different countries, addressing *RQ1*. Sections 4.3.1 and 4.3.2, respectively, examine the temporal distribution of multiple methods used by countries and the variety of methods employed over time, primarily addressing *RQ2*. Finally, Section 4.3.3 discusses the characteristics of research method development in different countries, addressing *RQ3*.

### 4.1 Distribution of articles from various countries

Figure 3 shows the cumulative percentage of articles in various countries, sorted by the number of articles. The statistics are presented for all authors, first authors, and corresponding authors, from left to right. We observe that the 20 countries with the highest number of articles, for all authors, first authors, or corresponding authors, account for over 90% of the total articles. The United States has the largest number of articles, representing 37.97% of all authors, 37.45% of first authors, and 40.76% of corresponding authors. This suggests that in international collaborations, more corresponding authors are from the United States.

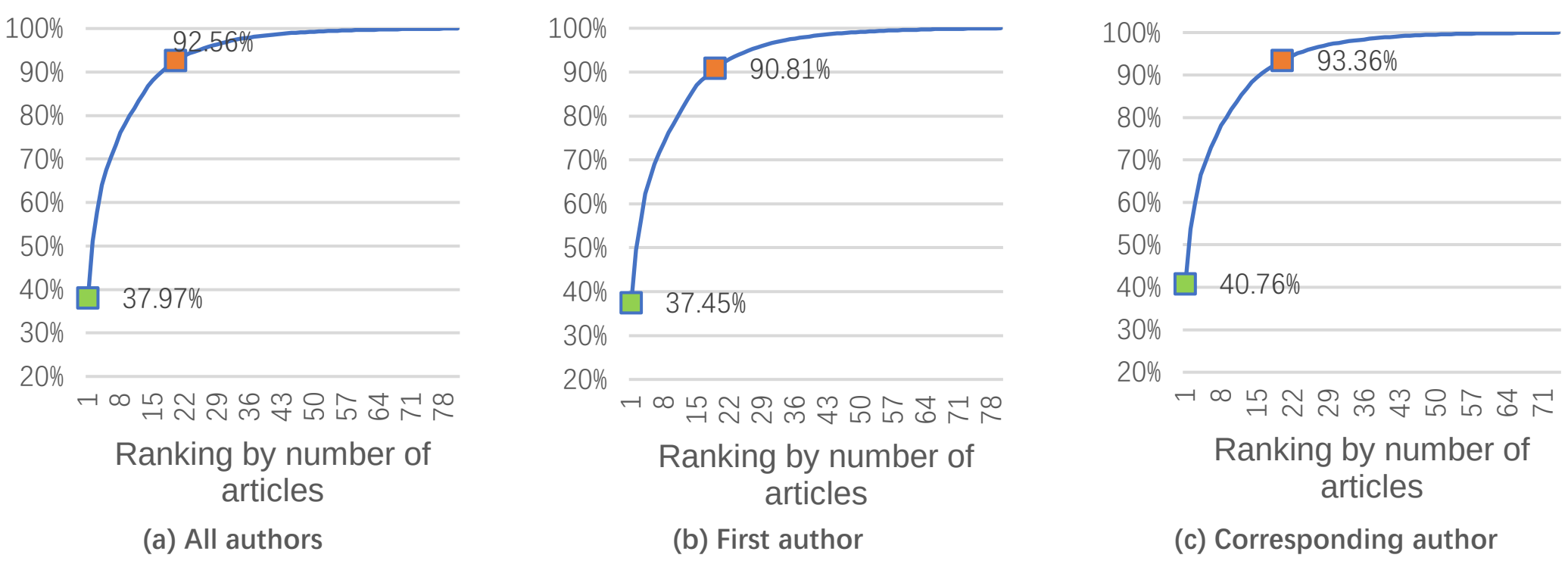

**Note:** The green squares in the chart represent the percentage of papers from the countries with the highest number of papers, while the orange squares represent the percentage of cumulative papers from the 20 most published countries.

**Figure 3. Cumulative percentage of articles by countries sorted by volume**

Figure 4 illustrates the distribution of articles among the top 20 countries with the highest publication counts. Notably, JASIST accounts for 67.78% of all papers, trailed by JDoc at 19.75%, and LISR at 12.45%. The figure demonstrates that various countries display a preference for publishing papers in distinct journals. For instance, the UK, Finland, Sweden, and Denmark exhibit a greater propensity to publish articles in JDoc. The red dot in Figure 4 represents each country's paper distribution across JASIST, JDoc, and LISR journals. It can be observed that the United States holds the largest share of published papers, succeeded by the United Kingdom, China, Canada, Australia, and others.

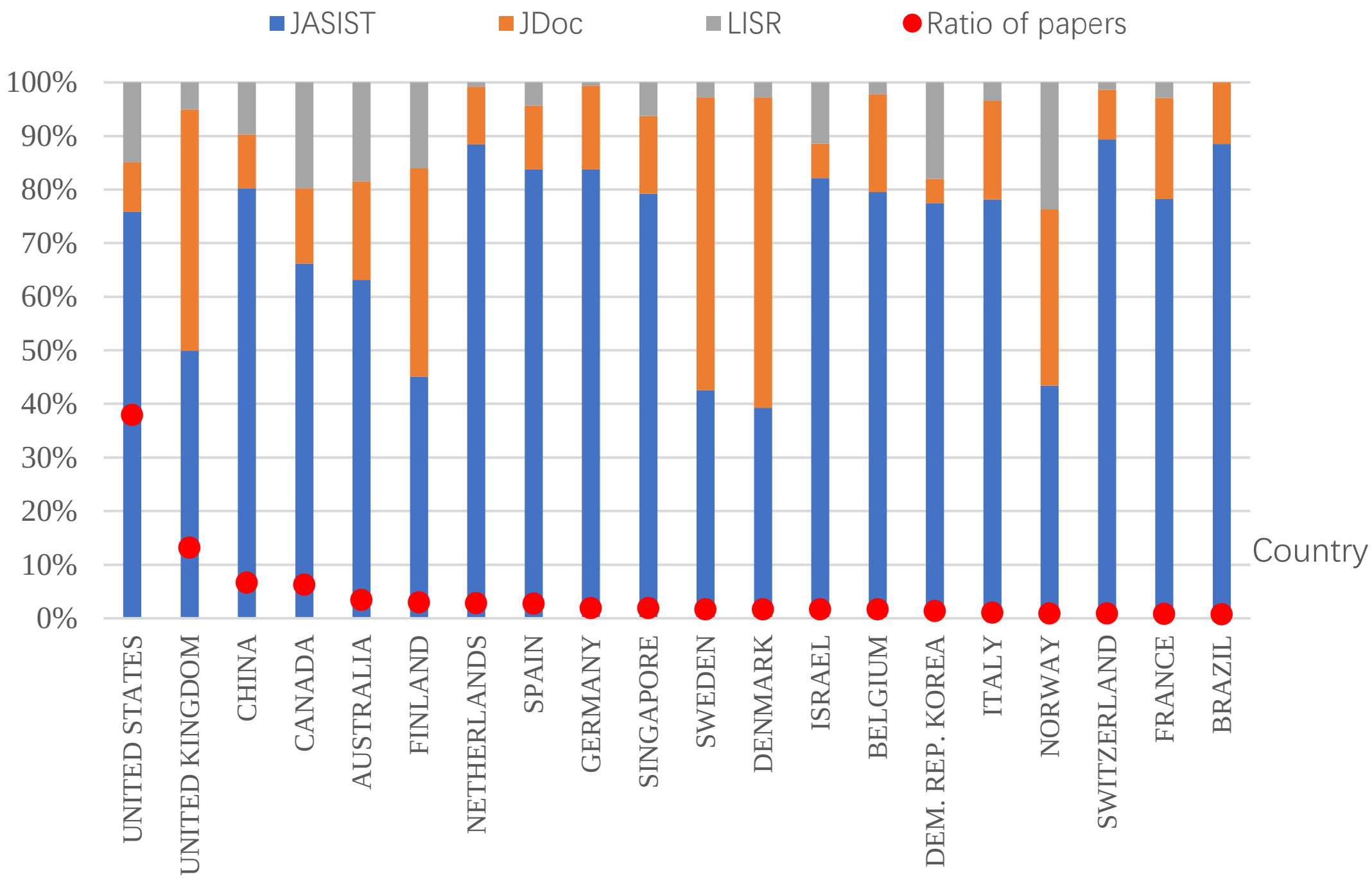

**Figure 4. Distribution of the 20 countries with the largest number of papers based on the statistics of all authors**

Each author's contribution in a paper is valuable. We believe that every author plays a distinct role. In the following analysis, we examine the countries of all authors of an article. Figure 5 presents the average number of countries involved in each paper from 1990 to 2019, showing a clear upward trend. This indicates an increase in international cooperation in Library and Information Science over the past 30 years

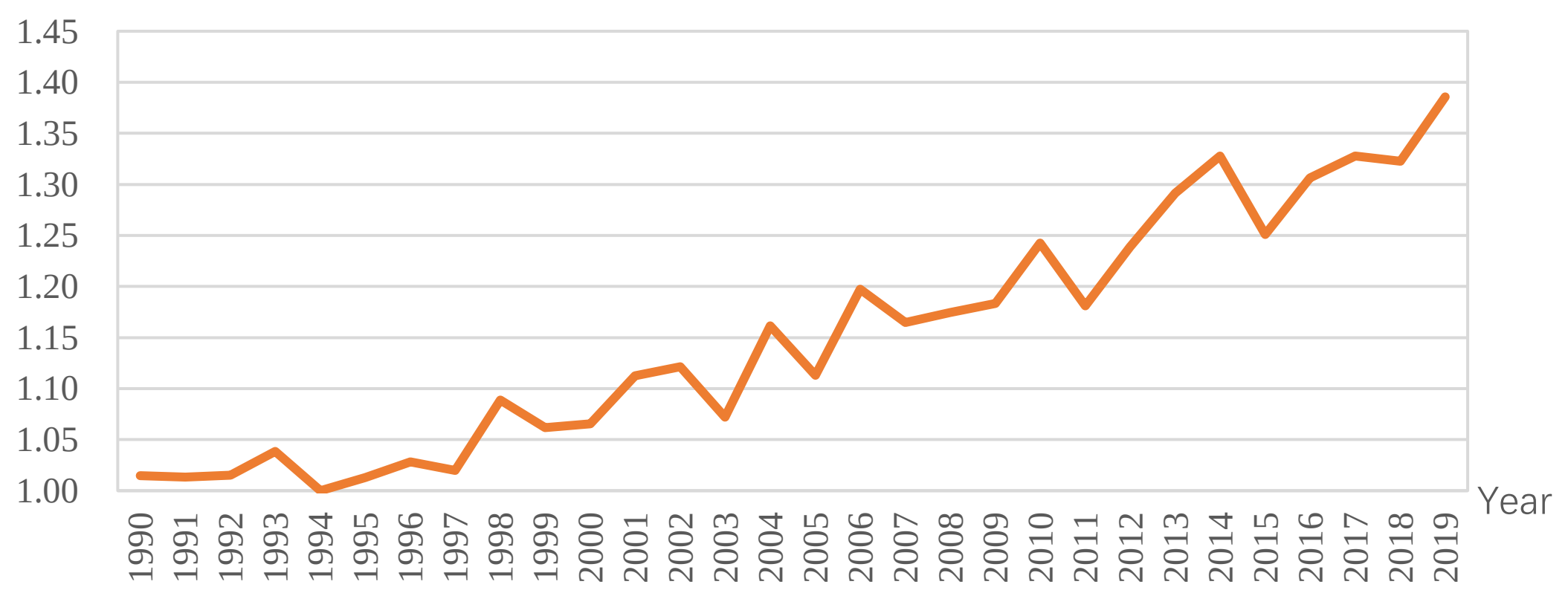

**Figure 5. Average number of countries involved in articles from 1990 to 2019**

### 4.2 Overall distribution of research methods used in different countries

This section aims to examine the overall distribution of research methods used in different countries. Firstly, we specifically analyze the most commonly used research methods across countries. Next, we visualize the degree of similarity in the usage of research methods across countries and investigate whether there are significant differences in their usage.

#### 4.2.1 The most used research methods in countries

We have noted that the top 20 countries, out of the 81 in total, contribute over 90% of the total article count in our dataset. Table 4 presents the five most commonly used research methods in the 20 countries with the highest number of research articles. The most commonly used research methods in each of the top 20 countries have been summarized and ranked. It has been shown that the most frequently used methods are experiment, bibliometrics, theoretical approach, and questionnaire. Additionally, if the top 5 most commonly used research methods in each country are summarized and ranked, the top 5 methods are content analysis, experiment, theoretical approach, questionnaire, and bibliometrics. Ma & Lund (2021) conducted a study using 3,422 articles published in LIS journals in the years 2006, 2012, and 2018 and found that the most popular research methods were experiment, questionnaire, and citation analysis/bibliographic data. The findings of both this study and Ma & Lund (2021) indicate that experiment and bibliometrics methods are widely used in research globally.

The most commonly used research methods in the Netherlands, Belgium, and Brazil account for over 40% of the total number of papers in these countries. In addition, China, Germany, Denmark, Israel, Italy, and France account for over 30% of the total number of papers in these countries. The application of bibliometrics reaches or even surpasses 50% in certain countries, such as the Netherlands, where 48.93% of papers utilize bibliometrics, and Belgium, with an outstanding rate of 60.58%. This highlights a significant preference for bibliometric research methods among scholars in these nations when contributing to JASIST, JDoc, and LISR journals. Within the leading 20 nations, it is solely in Norway that scholars predominantly employ questionnaire surveys as their principal research methodology in JASIST, JDoc, and LISR journals. In contrast, the remaining 19 countries exhibit a preference for research methods such as experiments, bibliometrics, and theoretical approaches.

Table 4 demonstrates that, among the top 20 nations, the proportion of research papers utilizing the five leading research methods makes up more than 60% of each country's total published papers in JASIST, JDoc, and LISR journals. Belgium has the highest proportion at 93.43%, while Sweden holds the lowest at 68.09%. Additionally, Table 4 reveals that certain countries, including Belgium, Switzerland, Italy, and Brazil, mainly use a limited number of research methods. In these countries, the top five methods listed in the table account for over 90% of the papers. Similarly, countries including Spain, the Netherlands, France, Germany, North Korea, China, Denmark, and Israel mostly use a few research methods, with the top five accounting for over 80% of the papers. The findings reveal that the top 20 countries mainly utilize a limited number of research methods in JASIST, JDoc, and LISR journals.

#### 4.2.2 Overall distribution of research methods across countries

Figures 6 present a two-dimensional visualization of the distribution of research methods across countries. To achieve this, we utilized the Uniform Manifold Approximation and Projection (UMAP) technique to project the high-dimensional research method distribution into two dimensions for visualization (McInnes *et al.*, 2018). Subsequently, we clustered the distribution of research methods across countries using a K-means algorithm (Wong & Hartigan, 1979) based on the Euclidean distance metric (Danielsson, 1980), and determined the number of clusters using the elbow method (Bholowalia & Kumar, 2014). In Figure 6, the closer the distance between countries, the more similar the distribution of research methods.

Figure 6 provides a visual representation of the distribution of research methods across countries, indicating which countries have a similar distribution of research methods. For instance, the research methods used by Brazil and the UK vary considerably, whereas countries such as Australia, New Zealand, and Canada exhibit a closer distribution of research methods. To determine whether the differences in research methods between countries were statistically significant, we conducted a Kolmogorov-Smirnov test (K-S test) and identified the countries with significant differences in the distribution of research methods, which are listed in Table 5.

Table 5 reveals that some countries exhibit significant differences in the usage of research methods, which can be divided into two categories: those between countries with a high volume of publications and those with a low volume of publications, and those between countries with a low volume of publications. Countries with a low volume of publications may focus on only a few research topics and use a limited number of research methods, resulting in a distribution of research methods that differs significantly from that of countries with a large number of publications. For instance, more than 60% of publications in BELGIUM employ “bibliometrics” as a research method. The differences in the distribution of research methods used among countries with fewer publications may reflect the various stages of development of the discipline in these two countries. For example, the most commonly used research methods in SINGAPORE are "experiment," "questionnaire," and "content analysis," while those in ITALY are "experiment," "questionnaire," and "content analysis".

**Table 4. The five most used methods in the 20 countries with the highest volume of papers**

| Country | Method 1 | % Papers used method 1 | Method 2 | % Papers used method 2 | Method 3 | % Papers used method 3 | Method 4 | % Papers used method 4 | Method 5 | % Papers used method4 | % Total |
|---|---|---|---|---|---|---|---|---|---|---|---|
| **UNITED STATES** | experiment | 19.62% | questionnaire | 15.06% | content analysis | 14.55% | theoretical approach | 13.01% | interview | 9.74% | 71.99% |
| **UNITED KINGDOM** | theoretical approach | 17.56% | interview | 13.49% | content analysis | 12.85% | experiment | 12.75% | questionnaire | 12.01% | 68.67% |
| **CHINA** | experiment | 32.42% | bibliometrics | 19.96% | questionnaire | 15.02% | content analysis | 8.79% | interview | 6.59% | 82.78% |
| **CANADA** | experiment | 15.15% | interview | 13.79% | theoretical approach | 13.40% | content analysis | 13.40% | questionnaire | 12.82% | 68.54% |
| **AUSTRALIA** | experiment | 18.79% | interview | 16.31% | theoretical approach | 14.18% | questionnaire | 13.83% | content analysis | 9.57% | 72.70% |
| **FINLAND** | theoretical approach | 17.21% | questionnaire | 17.21% | experiment | 15.98% | interview | 12.70% | content analysis | 12.70% | 75.82% |
| **NETHERLANDS** | bibliometrics | 48.93% | experiment | 15.02% | questionnaire | 8.58% | content analysis | 8.58% | theoretical approach | 5.15% | 86.27% |
| **SPAIN** | bibliometrics | 30.26% | experiment | 24.12% | theoretical approach | 14.04% | content analysis | 10.53% | webometrics | 7.46% | 86.40% |
| **GERMANY** | bibliometrics | 35.00% | experiment | 19.38% | content analysis | 11.25% | theoretical approach | 8.75% | questionnaire | 8.75% | 83.13% |
| **SINGAPORE** | experiment | 25.16% | questionnaire | 18.24% | content analysis | 17.61% | interview | 6.92% | theoretical approach | 5.66% | 73.58% |
| **SWEDEN** | theoretical approach | 17.02% | interview | 16.31% | bibliometrics | 14.89% | content analysis | 11.35% | observation | 8.51% | 68.09% |
| **DENMARK** | theoretical approach | 37.86% | bibliometrics | 21.43% | content analysis | 7.86% | experiment | 7.14% | questionnaire | 6.43% | 80.71% |
| **ISRAEL** | experiment | 30.71% | questionnaire | 20.00% | content analysis | 17.14% | webometrics | 6.43% | interview | 6.43% | 80.71% |
| **BELGIUM** | bibliometrics | 60.58% | theoretical approach | 18.25% | experiment | 5.84% | questionnaire | 4.38% | content analysis | 4.38% | 93.43% |
| **DEM. REP. KOREA** | experiment | 26.13% | questionnaire | 21.62% | content analysis | 15.32% | bibliometrics | 15.32% | transaction log analysis | 4.50% | 82.88% |
| **ITALY** | experiment | 32.18% | theoretical approach | 25.29% | bibliometrics | 21.84% | questionnaire | 6.90% | content analysis | 6.90% | 93.10% |
| **NORWAY** | **questionnaire** | **19.74%** | interview | 18.42% | theoretical approach | 17.11% | observation | 9.21% | content analysis | 9.21% | 73.68% |
| **SWITZERLAND** | bibliometrics | 28.00% | experiment | 22.67% | theoretical approach | 17.33% | questionnaire | 14.67% | content analysis | 10.67% | 93.33% |
| **FRANCE** | experiment | 34.78% | bibliometrics | 17.39% | theoretical approach | 14.49% | content analysis | 11.59% | interview | 5.80% | 84.06% |
| **BRAZIL** | experiment | 42.62% | theoretical approach | 19.67% | content analysis | 13.11% | bibliometrics | 11.48% | webometrics | 3.28% | 90.16% |

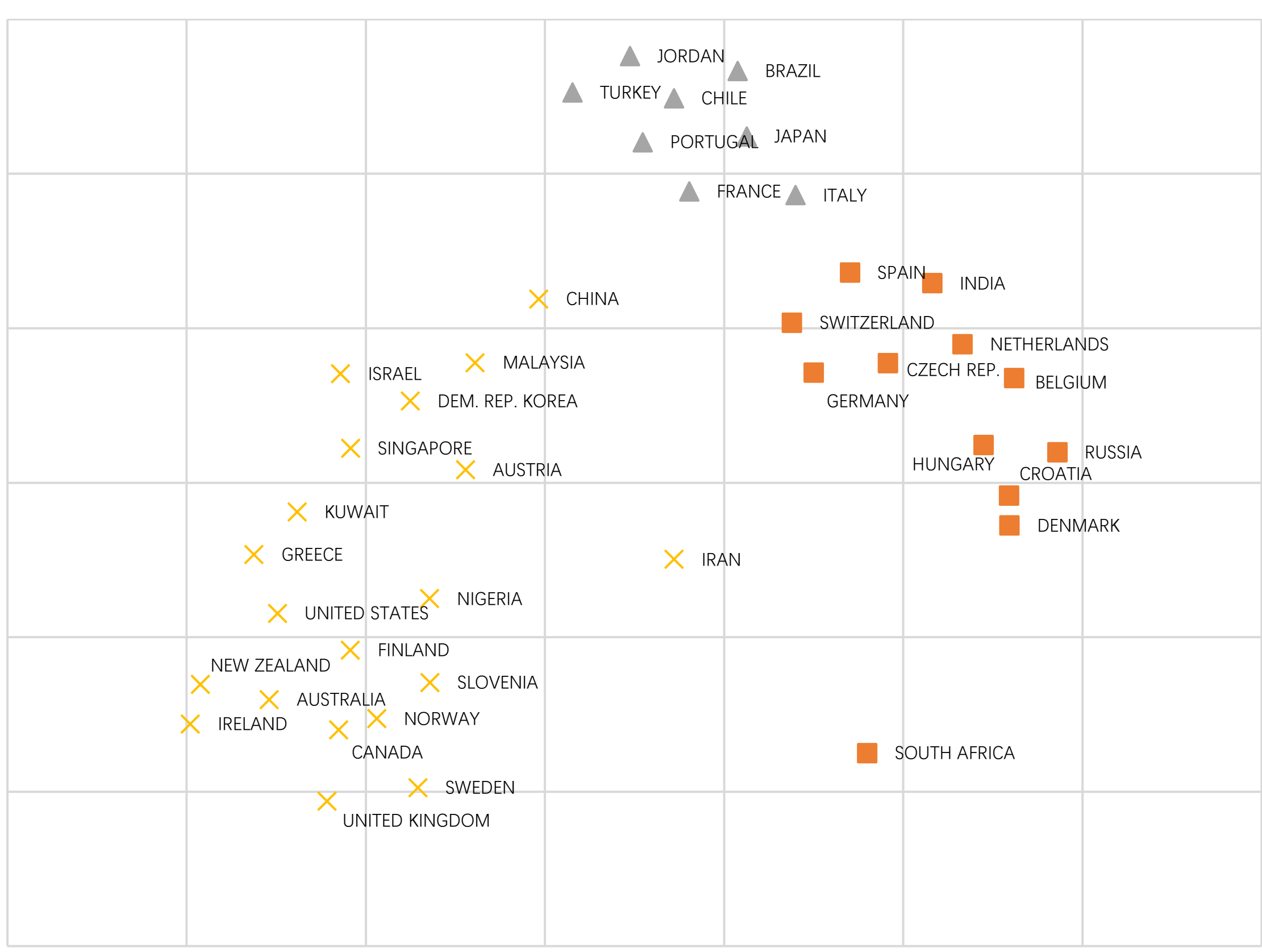

**Figure 6. Two-dimensional visualization of the usage of research methods by country**

**Table 5. Results of the K-S test for the distribution of research methods**

| Country | Country | K-S Statistic |
|---|---|---|
| UNITED STATES | ITALY | 0.56 |
| UNITED STATES | TURKEY | 0.63 |
| UNITED STATES | IRELAND | 0.56 |
| UNITED KINGDOM | BELGIUM | 0.50 |
| UNITED KINGDOM | ITALY | 0.50 |
| UNITED KINGDOM | TURKEY | 0.63 |
| UNITED KINGDOM | IRELAND | 0.56 |
| CHINA | TURKEY | 0.50 |
| CANADA | ITALY | 0.56 |
| CANADA | SWITZERLAND | 0.50 |
| CANADA | TURKEY | 0.63 |
| CANADA | IRELAND | 0.56 |
| AUSTRALIA | TURKEY | 0.56 |
| AUSTRALIA | IRELAND | 0.50 |
| GERMANY | TURKEY | 0.50 |
| SINGAPORE | ITALY | 0.50 |
| SINGAPORE | TURKEY | 0.56 |
| SINGAPORE | IRELAND | 0.50 |
| SWEDEN | TURKEY | 0.56 |
| SWEDEN | IRELAND | 0.50 |
| DENMARK | TURKEY | 0.50 |

**Note**: The critical values $D_{16,16,0.05}$ for the two-sample version of the K-S test is 0.48.

### 4.3 Distribution of research methods over time across countries

This section examines the temporal distribution of research methods across countries. We analyze changes in the use of research methods across countries from two perspectives: first, changes in the use of multiple methods across countries, and second, changes in the variety of research methods across countries. Finally, we cluster the trends based on the two types of change curves in research methods across countries.

#### 4.3.1 Time distribution of the multiple methods used by countries

We did not analyze research methods in terms of qualitative and quantitative, but rather by using multiple methods. This is because our method coding schema classifies research methods based on the data collection perspective. It is not reliable to directly define a research method as either qualitative or quantitative. Therefore, articles using a specific method can utilize either qualitative or quantitative methods. For instance, a questionnaire can be either qualitative or quantitative. Therefore, we need to examine the articles themselves to determine the type of method used, whether it is qualitative, quantitative, or mixed methods.

Figures 7 show the change in the percentage of the use of multiple methods in the three journals. As seen in the figures, the use of multiple methods has significantly increased over the 30-year period covered by our survey data. This finding is consistent with the results of previous studies (Chu, 2015; Hider & Pymm, 2008). Mixed research methods have been noticed and advocated as early as the early 20th century.

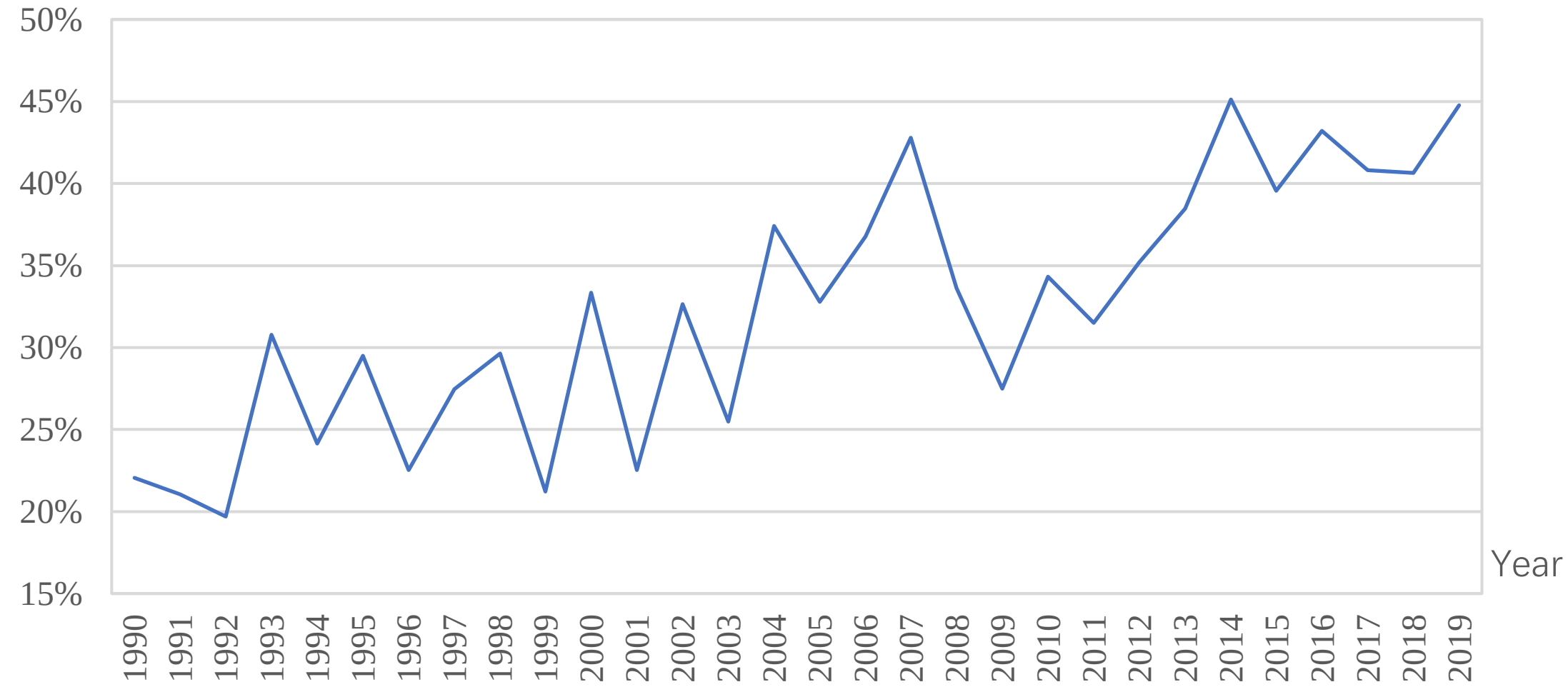

**Figure 7. Change in proportion of papers using multiple methods in LIS**

To further examine trends in the use of multiple methods in different countries, we also examined the use of multiple methods in each of the 20 countries with the highest number of publications. Figure 8 shows the distribution of the proportion of articles using multiple research methods in the top 20 countries. Some of the countries in the graph show large fluctuations in the curve due to the low number of publications in these countries in the early years. Looking at the variation across countries, we come to unexpected conclusions. We found that when looking at each country individually, the countries with the highest number of publications did not show a clear upward trend in using multiple research methods, which contradicts the overall results in Figure 8.

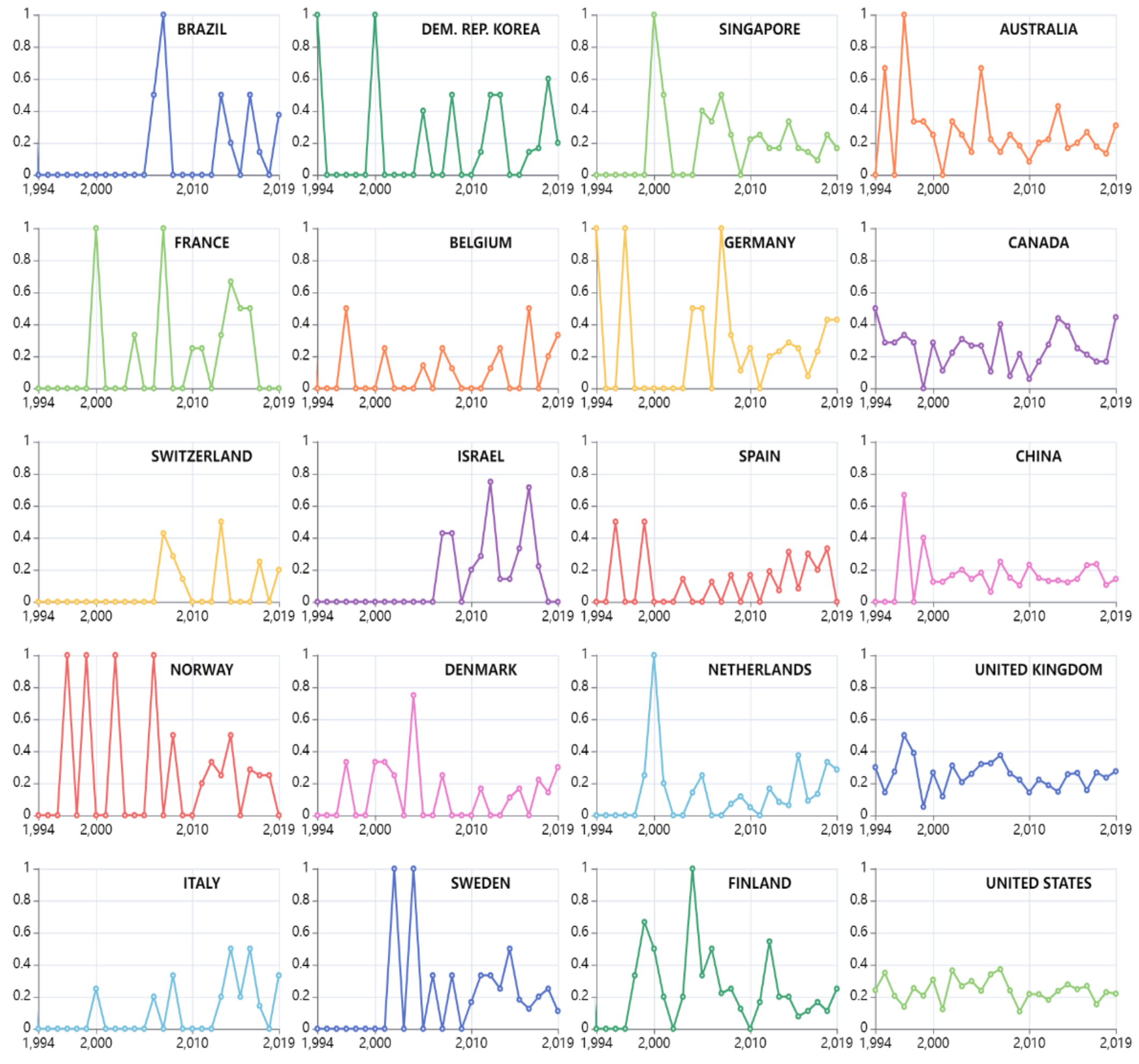

**Figure 8. Changes in the proportion of multiple research methods in the 20 most published countries**

The US and UK have the highest number of publications but do not show a clear trend toward an increase. In Figures 6-8, we find that there appears to be a greater bias towards using multiple methods in the early years when a country's publication volume increases in the three journals we examined. Typical examples include China, Australia, and France. We counted the use of multiple research methods in countries other than the UK and the US and compared them with the use of multiple research methods in the US and the UK. The results are shown in Figure 9. We believe that globally the use of multiple methods has indeed increased. But these increases have come more from countries with a lower volume of publications, and this has been particularly evident after 2008.

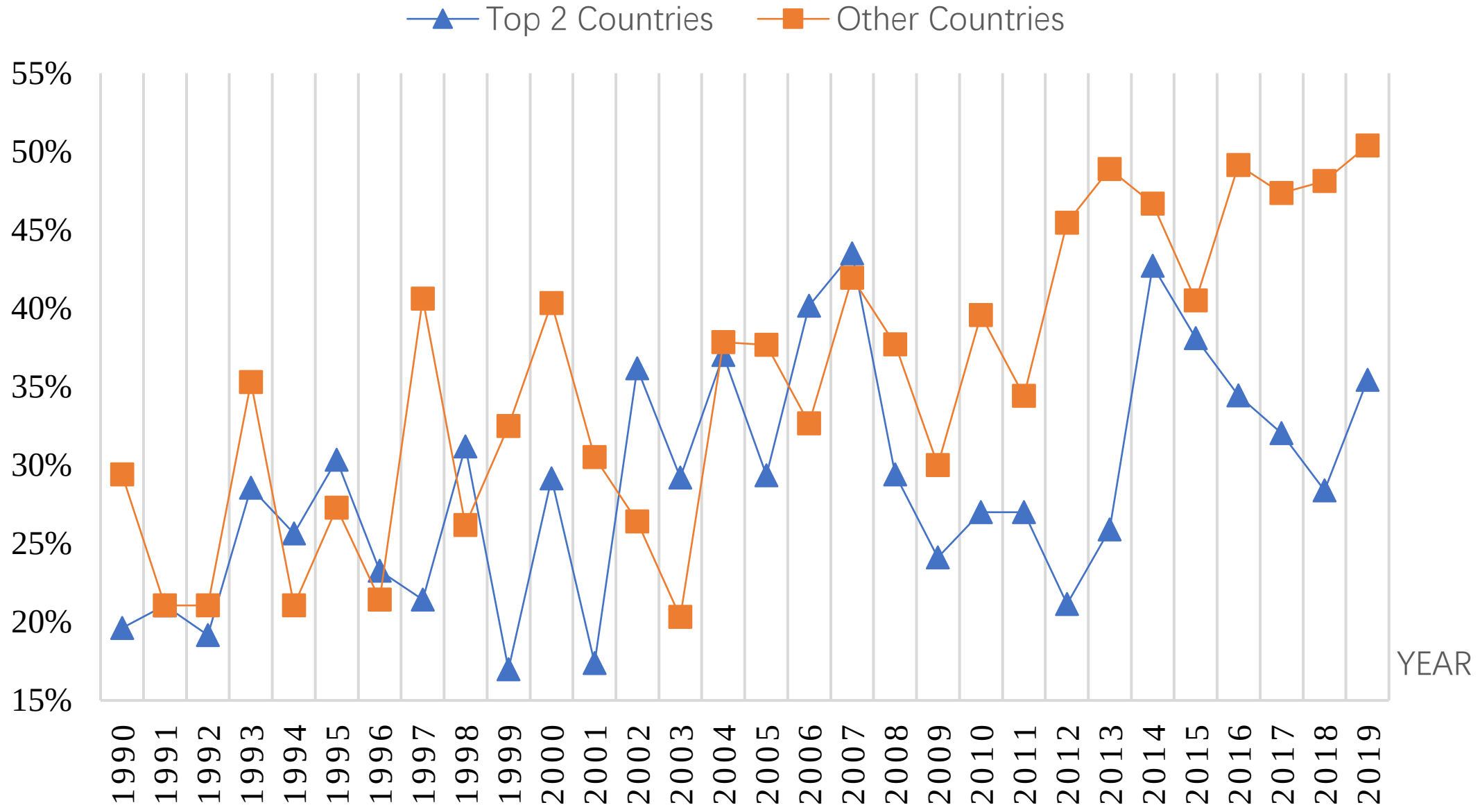

**Figure 9. Comparison of multiple method usage between the two countries with the highest publication volume and other countries**

### 4.3.2 Time distribution of the variety of methods used by countries

Figures 10 show the changes in the various research methods in the 20 countries with the highest number of articles. As we can see from the graph, most countries show an upward trend in the variety of research methods. Even the USA and the UK, where the discipline is at an advanced level of development, show a slight upward trend. Noteworthy here is Belgium, where the variety of research methods has always remained at a certain level, but there is no clear trend. As we can see from Table 4, the country makes extensive use of bibliometrics, a phenomenon that we believe is related to the country's disciplinary traditions.

The research questions determine the choice and use of research methods. Therefore, we analyzed the distribution of research topics across countries. We used Topic2vec (Angelov, 2020) to identify research topics for our survey data and obtained 17 research topics. After obtaining the research topics for each article, we calculated the trends in various research topics across countries, again based on Shannon Variety Index (Spellerberg & Fedor, 2003), and the results are shown in Figure 11. Figure 11 obtains similar results to Figure 10, for the obvious reason that the use of research methods depends on the research topic. An important reason for the increase in the variety of research methods in a country is the increase in the range of research topics in that country.

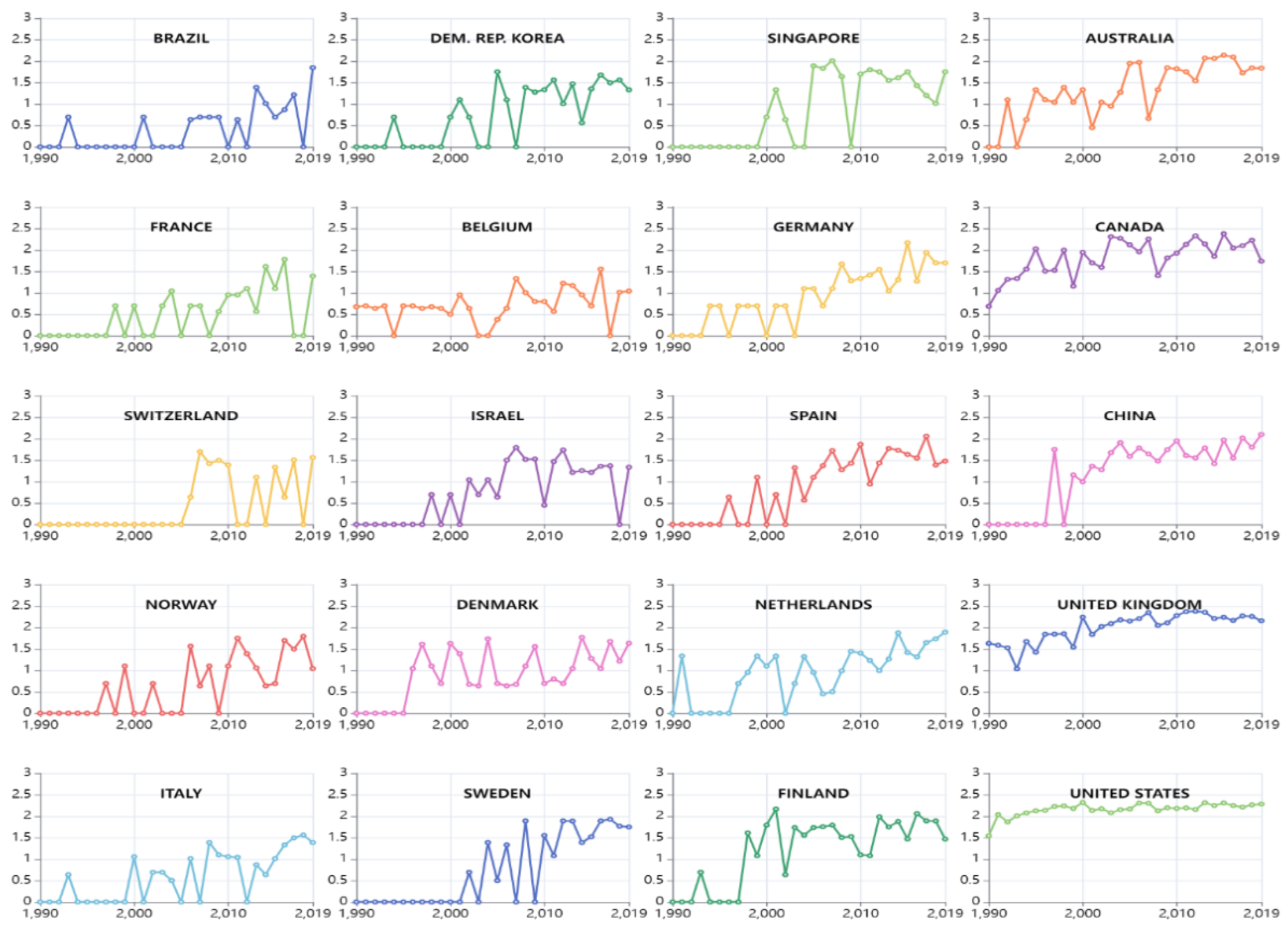

**Figure 10. Changes in variety of research methods among the 20 countries with the highest publication volume**

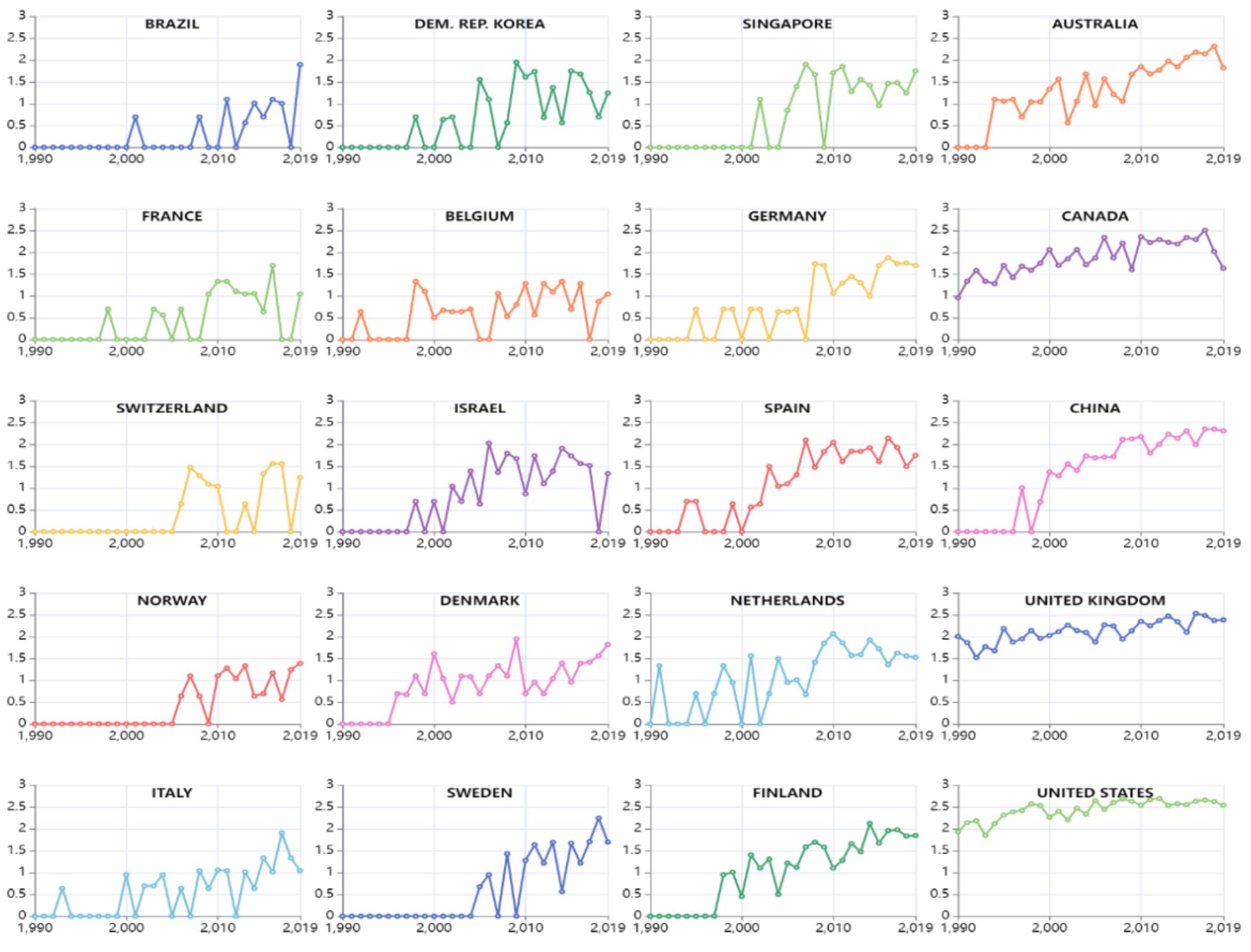

**Figure 11. Distribution of research topics among the 20 countries with the highest publication volume**

The use of research methods and research topics can reflect the state of disciplinary development in a country. Therefore, we aimed to cluster different countries based on the changes in the use of research methods in each country. The changes in the various research methods in different countries depicted in Figure 10 reflect the development of the discipline in various countries. We applied a dynamic time warping algorithm (Berndt & Clifford, 1994) to calculate the distance between the different curves and then clustered the curves in Figure 10 using a hierarchical clustering algorithm (Johnson, 1967). The clustering results are presented in Table 6.

**Table 6. Clustering results of method variety distribution curves for different countries**

| Cluster | Country |
|---|---|
| **1** | United States, United Kingdom, Canada |
| **2** | China, Australia, Finland, Netherlands, Spain, Germany, Singapore, Sweden, Denmark, Israel, Korea, Italy, Norway, Switzerland, France, Brazil |
| **3** | Belgium |

The clustering results are presented in Table 6, where the first cluster includes three countries that have maintained a high level of method variety consistently over the years, and we classify them as developed countries in the development of the LIS discipline. The second category includes 16 countries that show a clear upward trend in method variety over the last 30 years, and we consider them as countries where the LIS discipline has developed at a high rate. Some of these 16 countries, such as China, Australia, and Finland, have seen a marked increase in method variety over the last three decades. The third cluster includes only Belgium, and we believe that the development of the discipline in this country is somewhat specific and influenced by disciplinary traditions. The Netherlands have similar characteristics.

#### 4.3.3 Characteristics of the development of research methods in countries

The clusters shown in Table 6 indicate that there is little to no change in the variety of research methods in the first and third clusters. As a result, we will focus on the second cluster, which comprises countries with a rapid increase in the variety of research methods. In this section, we will analyze the characteristics of the development of research methods in these countries and explore which research methods are most used at different stages.

We attempted to classify the use and development of research methods in different countries over time using curve segmentation fitting. However, we could not obtain uniform-fitting results for different countries. Therefore, we divided the use of research methods in a country into five-year periods without dividing them into several phases.

Table 7 shows the most commonly used research methods in the countries with a rapidly increasing variety of research methods over time. We can see from Table 7 that different countries use different research methods at the same stage, indicating a clear difference in the direction of disciplinary development. Additionally, the research methods used at different stages in the same country also vary, which indicates differences in the development of the discipline at different times in the same country. These findings suggest that the development of the discipline varies from country to country based on their conditions, and there are no universal characteristics.

We aimed to investigate the differences between the distribution of research methods used in different countries on the same topic and the international distribution. To achieve this, we calculated the difference between the distribution of research methods and the international distribution for each country over ten years, for each research topic. To measure the differences between the distributions of research methods, we used the Jensen-Shannon divergence (Menéndez *et al.*, 1997).

**Table 7. The most used research methods in countries at different stages of rapid development**

| Country | 1990-1994 | 1995-1999 | 2000-2004 | 2005-2009 | 2010-2014 | 2015-2019 |
|---|---|---|---|---|---|---|
| China | theoretical approach | experiment | experiment | experiment | experiment | experiment |
| Australia | experiment | experiment | experiment | transaction log analysis | theoretical approach | interview |
| Finland | theoretical approach | interview | experiment | theoretical approach | content analysis | theoretical approach |
| Netherlands | bibliometrics | bibliometrics | bibliometrics | bibliometrics | bibliometrics | bibliometrics |
| Spain | theoretical approach | bibliometrics | experiment | theoretical approach | bibliometrics | experiment |
| Germany | experiment | bibliometrics | bibliometrics | experiment | bibliometrics | bibliometrics |
| Singapore | - | theoretical approach | experiment | experiment | experiment | content analysis |
| Sweden | bibliometrics | bibliometrics | interview | theoretical approach | interview | theoretical approach |
| Denmark | theoretical approach | theoretical approach | theoretical approach | bibliometrics | theoretical approach | theoretical approach |
| Israel | - | questionnaire | webometrics | experiment | experiment | experiment |
| Korea | experiment | experiment | questionnaire | experiment | questionnaire | experiment |
| Italy | theoretical approach | experiment | experiment | experiment | bibliometrics | theoretical approach |
| Norway | bibliometrics | interview | questionnaire | theoretical approach | questionnaire | interview |
| Switzerland | - | experiment | experiment | bibliometrics | bibliometrics | bibliometrics |
| France | questionnaire | bibliometrics | theoretical approach | experiment | experiment | experiment |
| Brazil | content analysis | - | experiment | experiment | experiment | experiment |

Although we obtained similar results for different research topics, we present only one example for clarity. In Figures 12, we show the variation in the distribution of research methods used by the top five countries with the highest number of publications and the international distribution for the "Information seeking behavior" topic. We found differences between countries and the international distribution, but these differences have decreased over the last three decades.

The UK and the US show the smallest differences between the international and national distributions. This is mainly due to their large share of publications. However, even these countries show a decreasing trend in the difference between the distribution of research methods used and the international distribution.

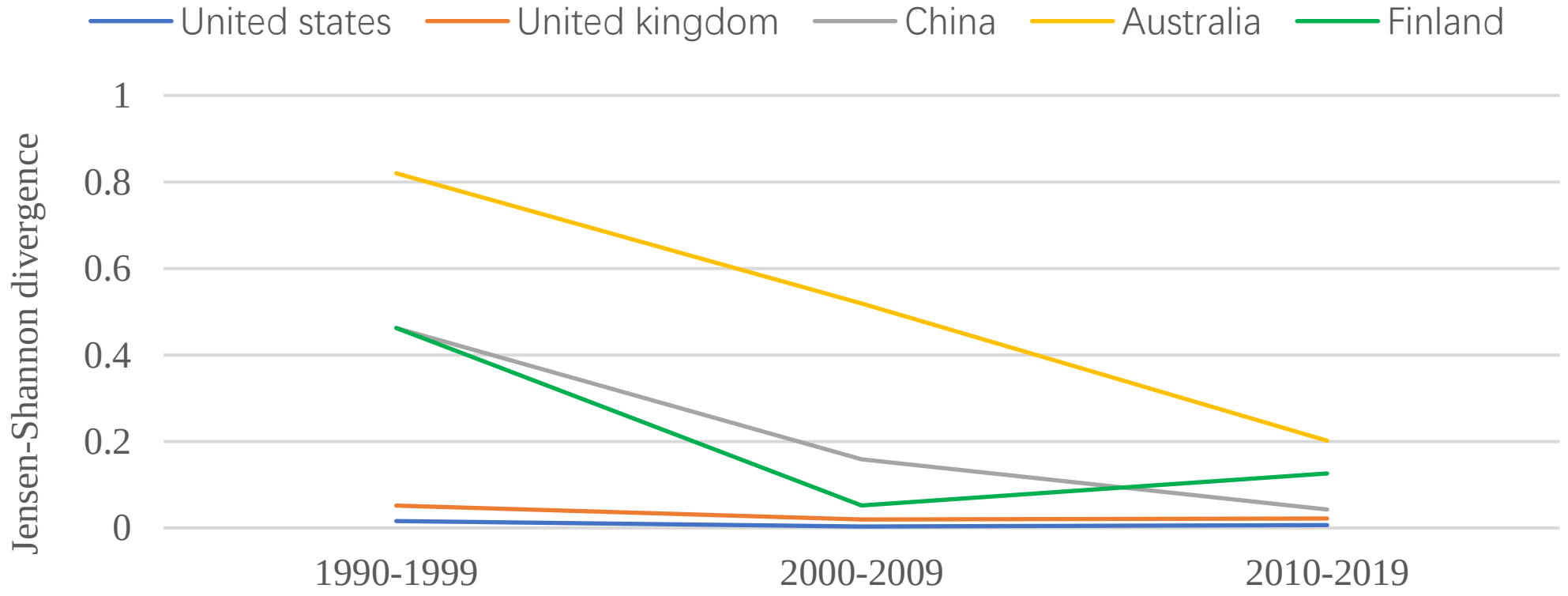

**Figure 12. The Jensen- divergence between the distribution of research methods used in “information behavior research” and international distribution for the top five countries**

## 5. Discussion and conclusion

The use of research methods is a crucial aspect of any discipline, including LIS. However, there is a lack of understanding regarding the differences in their usage across countries in the LIS field. To address this gap and gain a comprehensive understanding of the use of research methods in LIS globally, we conducted an automated coding and quantitative analysis of research methods in three core journals with international reach. Our analysis included data from 81 countries, and we focused on the 20 countries with the highest number of publications to examine their use of research methods.

### 5.1 International differences in LIS research

In our survey of international core journals, we found that the distribution of publications by country is highly concentrated, with the top twenty countries accounting for over 90% of the total number of articles. Moreover, our data shows that the average number of countries collaborating on each article has increased by 0.4 over the last thirty years, reflecting the significant rise in international collaboration in the field of Library and Information Science (LIS) over the past three decades.

Through quantitative analysis of the distribution of research methods across countries, as well as a specific examination of the research methods used in each country, we found significant differences in research method usage among some countries. For instance, Belgium and the Netherlands use bibliometric methods for over 45% of their total national publications. We also observed an increasing trend in research method variety across most countries in our data over the past thirty years, which is partly due to the expansion of research topics in each country.

Using cluster analysis, we categorized the changes in research methods among different countries over the last thirty years into three groups: developed countries, rapidly developing countries, and special countries with disciplinary traditions. We conclude that countries with disciplinary traditions have reached a relatively stable level of disciplinary development, while rapidly developing countries may progress into either developed or special countries with disciplinary traditions in the future, depending on various factors such as economic, social, and cultural aspects.

Our study also revealed that the use of multiple research methods has significantly increased in the LIS field over the last thirty years, which is consistent with previous studies (Fidel, 2008; Chu, 2015). However, we found that countries do not always maintain trends that align with international ones. After 2008, the increase in the use of multiple methods is more likely to come from rapidly developing countries with low publication volumes. Nevertheless, by comparing the distribution of research methods used in different countries within the same research topic and the international distribution, we found that differences between countries and international trends have decreased. This suggests that although there are still differences between countries in the LIS field, these differences are decreasing as research becomes more internationalized, and more countries are becoming integrated into national research collaborations.

### 5.2 Implications

The discussion in this paper presents a self-analysis of research methods used in the LIS field. An adequate understanding of research methods in LIS has practical implications for maintaining disciplinary independence, recognizing differences among countries, and guiding disciplinary development at the national level.

Compared to previous simple descriptive statistics (Lund & Wang, 2021; Ma & Lund, 2021), our study's quantitative approach makes the research methods used computable and comparable across countries, periods, and topics. This facilitates the analysis of trend evolution and quantifies the magnitude of differences. A broad understanding of the differences in research methods used in LIS in different countries can guide the search for transnational scientific cooperation and academic exchange.

This paper constructs a classification model based on full-text research methods. The automatic classification of research methods makes it easier to investigate the use of research methods within our discipline.

### 5.3 Limitations and feature work

The present research selected only three representative English journals in the LIS field to study. As we were examining the use of global research methods, we focused only on journals with high international impact. In our survey data, 81 countries have published in these journals. While these journals are internationally representative, examining the discipline's current state in a country should also consider important journals in the discipline from different countries separately. In our article, we attempted to reveal differences in the use of research methods in the LIS field in different countries, but we did not explain why such differences arise. The creation of differences in disciplinary development between countries is very complex and influenced by numerous economic and cultural factors. Discussing the methodological decisions of specific research is difficult and requires detailed research on specific cases to be feasible (Xie *et al.*, 2021).

In future research, we will use cross-lingual text processing techniques to assist in automatically encoding research methods for academic papers in different languages. To reveal the reasons for the differences in the use of research methods across countries, it should also be analyzed from the perspective of socio-cultural and economic development through survey methods. Although our data volume is not large enough, given that it is a global survey, we had to select journals with a long-term international impact. Future studies can expand the scope of the sample size and consider more journals from different countries to provide more comprehensive results. By combining quantitative and qualitative research methods, a more complete understanding of the use of research methods in the LIS field in different countries can be obtained.

### Acknowledgements

This work is supported by the National Natural Science Foundation of China (Grant No. 72074113). Thanks are due to Prof. Heting Chu for her valuable suggestions on this study.